\documentstyle[prd,aps,epsfig,floats]{revtex}
\begin{document}
%
\draft
\twocolumn[\hsize\textwidth\columnwidth\hsize\csname
@twocolumnfalse\endcsname
\title{
Effect of Fourth Generation $CP$ Phase on $b\to s$ Transitions}
\author{Abdesslam Arhrib$^{1,2}$ and Wei-Shu Hou$^3$\\}
\address{
1: Max-Planck Institut f\"ur Physik, F\"ohringer Ring 6,
D-80805 M\"unchen, Germany \\
2: LPHEA, Physics Department, Faculty of Science-Semlalia,
P.O.B 2390, Marrakesh, Morocco\\
3: Department of Physics, National Taiwan University, Taipei,
Taiwan 10764, R.O.C. }
\date{\today}
\maketitle

\begin{abstract}
We study the effect of a sequential fourth generation on 
$b\to s\gamma$, $B_s$ mixing, 
$B\to K^{(*)}\ell^+\ell^-$, $X_s\ell^+\ell^-$ and $\phi K_S$, 
taking into account the presence of a new $CP$ phase in quark mixing. 
We find that the effects via electromagnetic and strong penguins, 
such as in $b\to s\gamma$ or 
$CP$ violation in $B\to \phi K_S$, are rather mild, 
but impact via box or $Z$ penguins is quite significant.
Not only can one have much larger than expected $B_s$ mixing,
even if it is measured near the present limit,
mixing dependent $CP$ violation could become maximal.
For electroweak penguin modes, the fourth generation can
provide the enhancement that perhaps may be needed,
while the $m_{\ell^+\ell^-}$ dependence of 
the forward-backward asymmetry in $B\to K^{*}\ell^+\ell^-$
can further probe for the fourth generation.
These effects can be studied in detail at the B factories 
and the Tevatron.
\end{abstract}
]

\section{Introduction}

The source of $CP$ violation within the Standard Model lies
in the Cabibbo-Kobayashi-Maskawa (CKM) quark mixing matrix, where
a unique KM phase emerges with 3 generations of quarks and
leptons. Despite its great success, there are reasons to believe
that the 3 generation Standard Model (SM, or SM3) may be incomplete.
For example, the generation structure is not understood,
while recent observations of neutrino oscillations point towards
an enlarged neutrino sector~\cite{neutrino}.
A simple enlargement is to add a sequential fourth generation (SM4);
with additional mixing elements, one crucial aspect is that
the source for $CP$ violation is no longer unique.

The B factories at KEK and SLAC have turned on with a splash, and
$CP$ violation has been observed in $B^0\to J/\psi K_S$ decay
modes that is consistent with SM3. Both experiments have recently
observed the $B\to K\ell^+\ell^-$~\cite{belle,belle2,babar,babar2}
decay mode, while the Belle Collaboration has just reported the
observation of the inclusive $B\to X_s\ell^+ \ell^-$
mode~\cite{belleXsll}. Though a bit on the high side, the results
are again not inconsistent with SM3. However, there may be a hint
for new physics in the mixing dependent $CP$ violation in $B^0\to
\phi K_S$ decay~\cite{phiKs}. At any rate, together with many
related loop-induced processes, we have entered an age where the
uniqueness of the 3 generation SM phase can start to be checked.

The $B\to K\ell^+ \ell^-$ type of decays are
highly suppressed in the SM because they occur only via
electroweak penguin processes. The first measurement of such
processes through $b\to s\gamma$ was performed some years ago by
CLEO~\cite{cleo95}.  The measured branching ratio has been used to
constrain the Wilson coefficient $\vert C_{7}^{\rm eff}\vert$
\cite{Borzumati}. The measurement of $B\to (K,\ K^*)\ell^+ \ell^-$
and $X_s\ell^+ \ell^-$ modes can provide information on
$C_{9}^{\rm eff}$ and $C_{10}$, as well as sign information on
$C_{7}^{\rm eff}$.
The contributions from fourth generation to rare decays have been
extensively studied~\cite{Hou,wshou,fourth,Huo}, where the
measured $b\to s\gamma$ decay rate has been used~\cite{Huo} to put
stringent constraints on the additional CKM matrix elements.
However, most of these studies neglect the effect of new $CP$
phases by taking $\lambda_{t'} \equiv V_{t's}^*V_{t'b}$ to be
real. The $b\to s\gamma$ constraint then implies that, for fixed
$t'$ quark mass, $\lambda_{t'}$ can take on only one of two
values.

In this paper, we study the effect of the fourth generation on
$b\to s$ transitions by including the $CP$ phase in
$V_{t's}^*V_{t'b}$. We use the updated measured value of $b\to s
\gamma$~\cite{PDG} and the lower bound on $B^0_s$--$\bar B^0_s$
mixing, to put constraint on fourth generation parameters. The
effect of the allowed region of parameter space on $b\to s$
transitions are compared with Belle and BaBar measurements. 
We further discuss the 
forward-backward asymmetry of $B\to (K,\ K^*)\ell^+ \ell^-$, and
indirect $CP$ violation in $B_s$-$\bar B_s$ mixings and $B^0\to
\phi K_S$ decay. We stress that we focus on $V_{t's}^*V_{t'b}$
only, in anticipation of new data. Thus, only one additional new
$CP$ phase is introduced. The $K^+\to\pi^+\nu\bar\nu$ process,
$B_d$ mixing and its $CP$ phase etc., involve $V_{t'd}^*V_{t's}$,
$V_{t'd}^*V_{t'b}$ respectively, which are quite independent from
the $CP$ phase information contained in $V_{t's}V_{t'b}$.

\section{Strategy}

The $b\to s\gamma$ process provides a stringent constraint on new
physics, while the nonobservation of $B_s$--$\bar B_s$ mixing also
puts constraints on a fourth generation. We shall discuss these
two traditional constraints before turning to more recent
experimental observations.

\subsection{The $b\to s\gamma$ Constraint}

With a sequential fourth generation, the Wilson coefficients $C_7$
and $C_8$ receive contributions from $t'$ quark loop, which we
will denote as $C_{7,8}^{\rm new}$. Because a sequential fourth
generation couples in a similar way to the photon and W, the
effective Hamiltonian relevant for $b\to s\gamma$ decay has the
following form:
\begin{eqnarray}
{\cal H}_{\rm eff} = \frac{4 G_F}{\sqrt{2}} \sum_{i=1}^{i=8}
[\lambda_{t} C_i^{\rm SM}(\mu) + \lambda_{t'} C_i^{\rm new}(\mu) ]
O_i(\mu),
\label{bsg}
\end{eqnarray}
with $\lambda_f=V_{fs}^* V_{fb}$, and the $O_i$s are just as in
SM. The unitarity of the $4\times 4$ CKM matrix leads to:
$\lambda_u + \lambda_c + \lambda_t + \lambda_{t'}=0$. Taking into
account that $\lambda_u$ is very small compared to the others, one
has~\cite{Hou,wshou}:
\begin{eqnarray}
\lambda_t\cong-\lambda_c - \lambda_{t'}, \label{uni4}
\end{eqnarray}
and $\lambda_c = V_{cs}^*V_{cb}$ is real by convention.
It follows that
\begin{eqnarray}
\lambda_t C_{7,8}^{\rm SM}+\lambda_{t^\prime}C_{7,8}^{\rm new} =
-\lambda_c C_{7,8}^{\rm SM} +\lambda_{t'} (C_{7,8}^{\rm new}
-C_{7,8}^{SM}), \label{decomp}
\end{eqnarray}
where the first term corresponds to the usual SM contribution. It
is clear that, for $m_{t'}\to m_t$ and $\lambda_{t'} \to 0$
limits, the $\lambda_{t'} (C_{7,8}^{\rm new} -C_{7,8}^{SM})$ term
vanishes, as required by the GIM mechanism.  
We parameterize 
\begin{eqnarray}
\lambda_{t'} \equiv r_s \, e^{i \Phi_s},
 \label{ltp}
\end{eqnarray}
where $\Phi_s$ is a new CP violating phase.

The branching ratio of $b\to s\gamma$ is computed by using
formulas given in \cite{buras}. Using ${\cal B}(B\to X_c
e\bar{\nu}_e)=10.4\% $, $\lambda_c=0.04$, we find ${\cal B}(B\to
X_s \gamma)\simeq 3.26\times 10^{-4}$ in SM. The present world
average for $B\to X_s \gamma$ rate is \cite{PDG},
\begin{eqnarray}
{\cal B}(B\to X_s \gamma) = (3.3\pm 0.4)\times 10^{-4}.
 \label{bsA}
\end{eqnarray}
We will keep the $b\to s\gamma$ branching ratio in the 1$\sigma$
range of (2.9--3.7) $\times 10^{-4}$ in the presence of fourth
generation.

We note that the CDF collaboration excludes \cite{cdf4} the $b'$
quark in the mass range of 100 GeV $< m_{b'}< 199 $ GeV at 95\% C.L.,
if ${\cal B}(b'\to b Z)=100\%$.~\footnote{Such limit can be
relaxed \cite{AH} if we consider that the decays $b'\to b H$ and
$b'\to c W$ may be competitive with $b'\to b Z$.}
Precise EW data provide stringent constraint on the fourth
generation: the splitting between $t'$ and $b'$ is severely
constrained by $|m_{t'}-m_{b'}|\leq m_Z$ \cite{PDG,AH,sher}.
Assuming that $m_{t'}>m_{b'}$, we conclude that the $t'$ should be
heavier than $\approx 199$ GeV $+ m_Z$. In our analysis we will
consider only $t'$ heavier than 250 GeV.

\begin{figure}[b!]
\centerline{{\epsfxsize1.6 in \epsffile{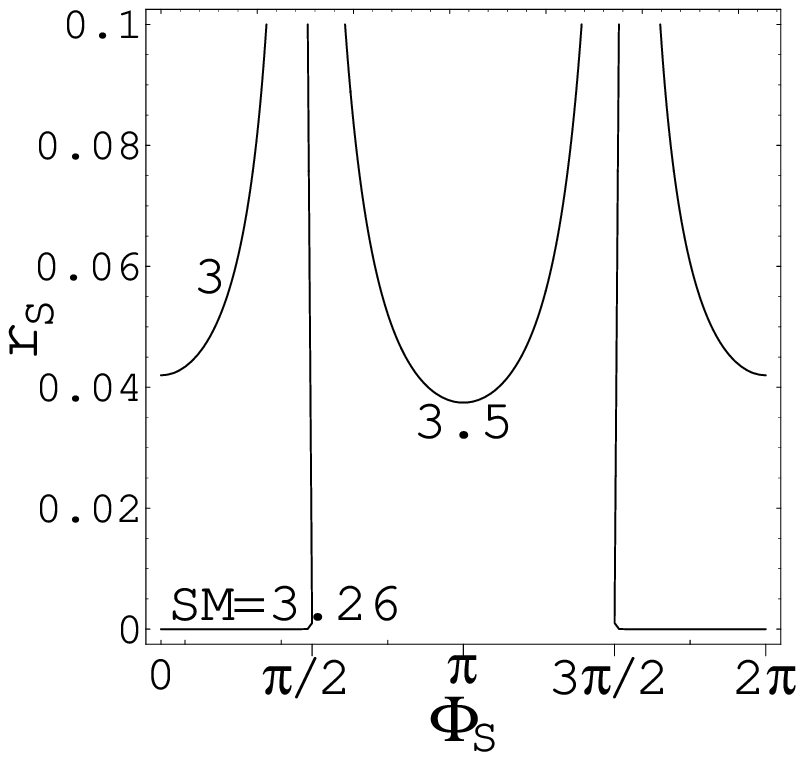}}
 {\epsfxsize1.6 in \epsffile{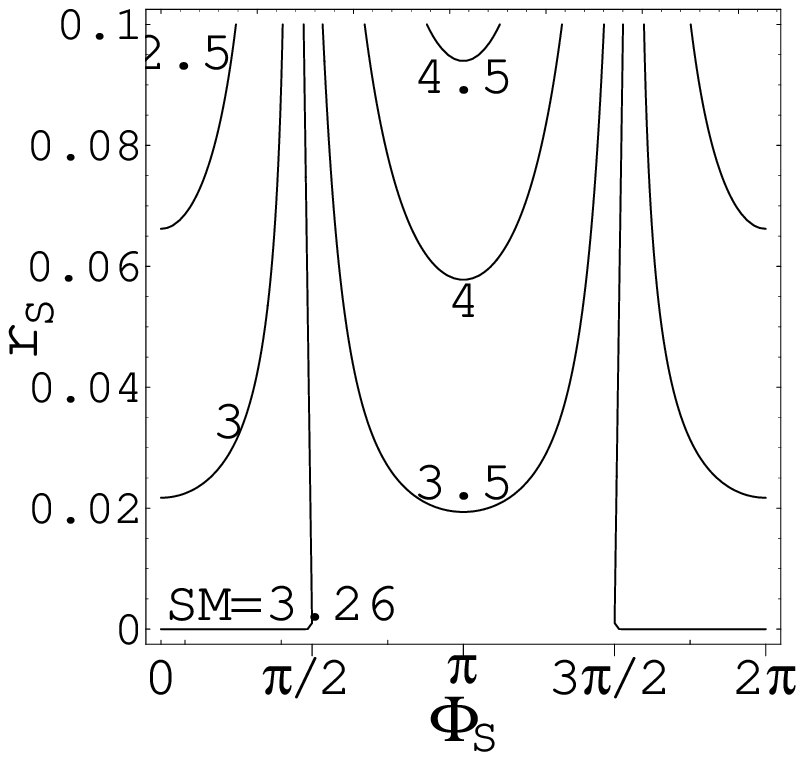}}}
\smallskip
\caption{Contours for ${\cal B}(B \to X_s \gamma)(\times 10^4)$ in
the $(\Phi_s,r_s)$ plane, for $m_{t^{\prime}}=250$ (350) GeV on
the left (right). } \label{fig1}
\end{figure}

In Fig.~\ref{fig1} we give a contour plot of ${\cal B} (B\to X_s
\gamma)$ in $(\Phi_s, r_s)$ plane for $m_{t'}=250$ and 350 GeV. We
see that the $CP$ phase $\Phi_s$ does make considerable impact on
the range of ${\cal B} (B\to X_s \gamma)$, but one is not
particularly sensitive to $m_{t'}$. In general, for both
$m_{t'}=250$ and 350 GeV cases, $r_s$ cannot be much larger than
0.02 if one wishes to keep ${\cal B}(B\to X_s \gamma)$ in the
range of (2.9--3.7) $\times 10^{-4}$. But for $\Phi_s \sim \pi/2$,
$3\pi/2$, {\it when $V_{t's}^*V_{t'b}$ is largely imaginary}, $r_s$ can
take on much larger values; for these cases the $t^\prime$
contribution adds only in quadrature to SM contribution, hence
much more accommodating. In this sense, our approach is different
from the approach of \cite{Hou,Huo} where $\lambda_{t'}$ is taken
as real, and can take on only one of two values for a fixed
$m_{t'}$.

Since the chiral structure is the same as in SM, 
and since a heavy state such as $t'$ does not bring in 
extra absorptive parts,
$CP$ asymmetry in $b\to s\gamma$ remains small.
However, the possibility of sizable $CP$ odd $t'$ contributions 
allowed by $b\to s\gamma$, while not easily
distinguished in the $b\to s\gamma$ process itself, can make
impact on $CP$ observables in other $b\to s$ transitions, as we
shall see.

\subsection{$B_s^0$--$\bar B_s^0$ Mixing Constraint}

Because of strong $m_{t'}$ dependence, the fourth generation
contribution can easily impact on $B^0_s$--$\bar B^0_s$ mixing and
its $CP$ phase $\Phi_{B_s}$, and can be accessible soon at the
Tevatron Run II.

We use the definition of $\Delta m_{B_s} = 2 |M^B_{12}|$ and
$M^B_{12}=|M^B_{12}|e^{i\Phi_{B_S}}$, where $M^B_{12}$ is given by
\cite{inami}:
\begin{eqnarray}
M^B_{12} &=&\frac{G_F^2 M_W^2 }{12 \pi^2} M_{B_s}
{B_{B_s}}f_{B_s}^2 \{ \lambda_t^2 \eta S_0(x_{tW}^2) \nonumber \\
&&+\eta^\prime \lambda_{t'}^2 S_0(x_{t'W}^2) + 2
\widetilde{\eta}^\prime\ \lambda_t\ \lambda_{t'}\
\widetilde{S}_0(x_{tW}^2,x_{t'W}^2) \}, \label{m12}
\end{eqnarray}
with $x_{fW}=\frac{m_f}{M_W}$, and
\begin{eqnarray}
S_0(x)& = & \frac{4 x -11 x^2 +x^3 }{4(1-x)^2 }-\frac{3}{2}
\frac{\log x x^3}{(1-x)^3}, \\
\frac{\widetilde{S}_0(x,y)}{x y} & = &  \left\{
\frac{1}{y-x}\left[\frac{1}{4}+
\frac{3}{2}\frac{1}{1-y}-\frac{3}{4}
\frac{1}{(1-y)^2}\right] \log y\right. \nonumber \\
 & + &  \frac{1}{x-y}
\left[\frac{1}{4}+\frac{3}{2}\frac{1}{(1-x)}-
\frac{3}{4}\frac{1}{(1-x)^2}\right]\log x
\nonumber \\
 & - & \left. \frac{3}{4} \frac{1}{(1-x)(1-y)}
\right\},
\end{eqnarray}
where $\eta=0.55$ is the QCD correction factor. Taking into
account the threshold effect from $b'$ quark, we find
$\eta^\prime=\alpha_s(m_t)^{\frac{6}{23}} (\frac{\alpha_s(m_{b'})
}{\alpha_s(m_t ) } )^{\frac{6}{21}} (\frac{\alpha_s(m_{t'})
}{\alpha_s(m_{b'} ) } )^{\frac{6}{19}}$. Given the splitting
between $t'$ and $b'$, it turns out that $\eta^\prime$ is also
close to 0.55 for  250 GeV $< m_{t'} < 350 $ GeV, which is not
surprising. We shall take $\widetilde{\eta}^\prime = \eta^\prime$.

\begin{figure}[t!]
\centerline{{\epsfxsize1.6 in \epsffile{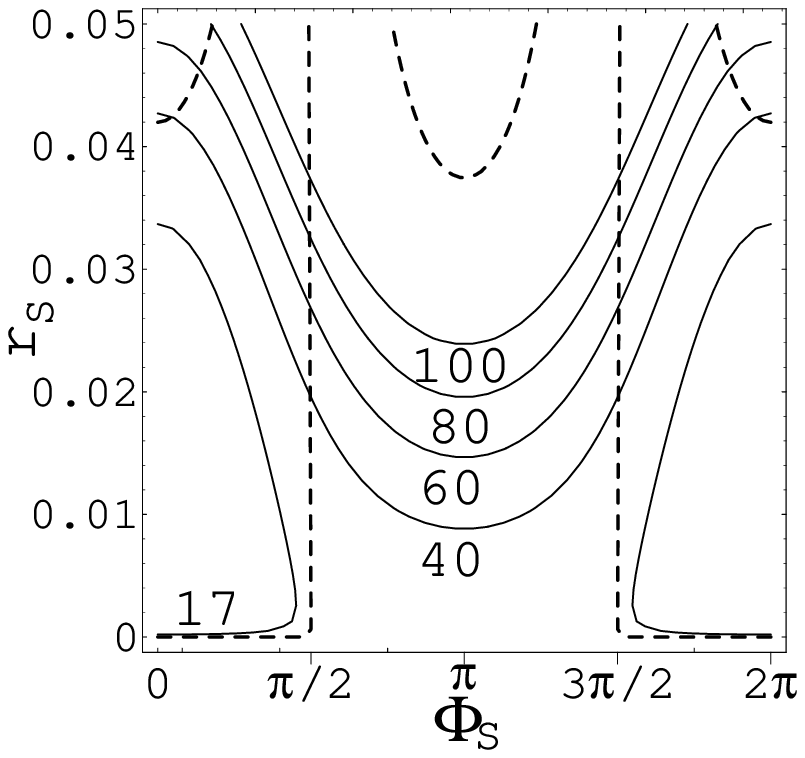}}
 {\epsfxsize1.6 in \epsffile{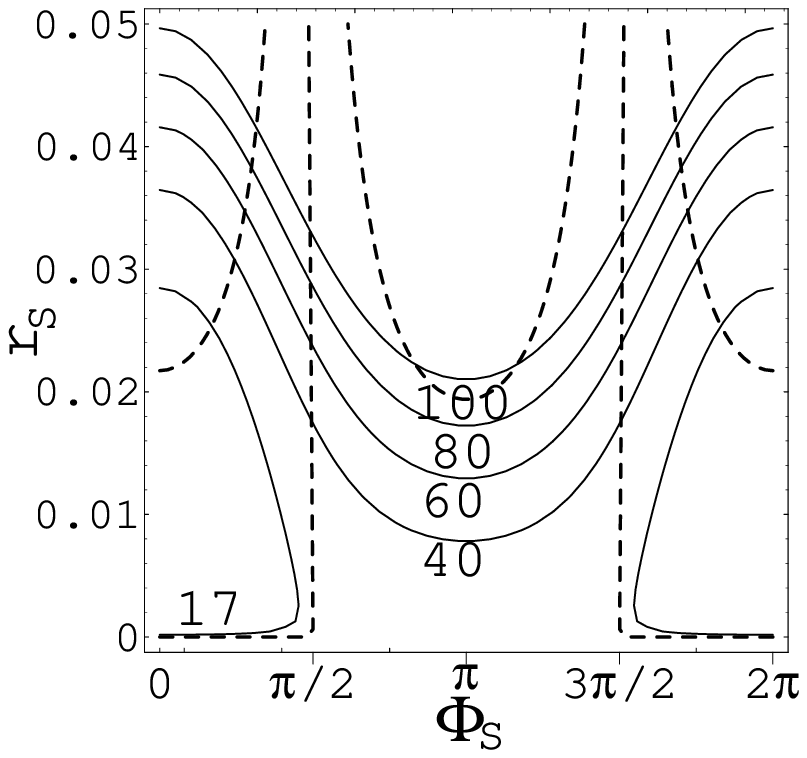}}}
\smallskip
\caption{Contours for $\Delta m_{B_s}$ (ps$^{-1}$) in the
$(\Phi_s,r_s)$ plane, for $m_{t^{\prime}}=250$ (350) GeV on the
left (right), with the results of Fig. 1 overlaid as dashed lines
for comparison.} \label{newfig3}
\end{figure}

In our calculation we use $B_{B_s} f_{B_s}^2=(260\ {\rm MeV})^2$
and $\overline{m}_t(m_t)=168$ GeV, which gives $\Delta
m_{B_s}^{\rm SM}\simeq 17.0\ {\rm ps}^{-1}$ with vanishing $\sin
2\Phi_{B_s}^{\rm SM}$. For illustration, in Fig.~\ref{newfig3} we
give a contour plot of $\Delta m_{B_s}$ in $\Phi_s$--$r_s$ plane,
for $m_{t'}=250$ and $350$ GeV. The result of
Fig.~\ref{fig1} is overlaid for comparison. We see that $\Delta
m_{B_s}$ can reach 90 ${\rm ps}^{-1}$ for $r_s\approx 0.02$ and
$\Phi_s\approx \pi$, but for heavier $m_{t'}$, ${\cal B}(B\to X_s
\gamma)$ will reach beyond $3.7\times 10^{-4}$ and become less
likely. For $\Phi_s \sim \pi/2$, $3\pi/2$, where $b\to s\gamma$ is
more tolerant, $\Delta m_{B_s}$ can get greatly enhanced. On the
other hand, the experimental lower bound of $\Delta m_{B_s}>14.9\
{\rm ps}^{-1}$ implies that the region where $0\leq r_s\leq 0.03$
and $\cos\Phi_s>0$ is disfavored, which rules out a region allowed
by $b\to s\gamma$. We now see that the allowed parameter space is
larger for the $m_{t'} = 250$ GeV case, i.e. for heavier $m_{t'}$,
$r_s$ can only take on smaller values. This is because the
$m_{t'}$ dependence for $\Delta m_{B_s}$ is rather strong, but
is much weaker in case of $b\to s\gamma$.

We stress that the $\Phi_s \sim \pi/2$, $3\pi/2$ cases are much
more forgiving because the $CP$ phase for $b\to s$ transitions is
practically real in SM3, hence a  purely imaginary fourth
generation contribution adds only in quadrature. We thus turn to
study the possible impact of a fourth generation, especially
through its new $CP$ phase, on other $b\to s$ transitions.

\section{$CP$ Violation in $B$ Meson Mixing}

Here we mean $B_s^0$--$\bar B_s^0$ mixing. A fourth generation can
affect $B_d^0$--$\bar B_d^0$ mixing and its $CP$ phase through
$V_{t'd}^*V_{t'b}$. Since there is no apparent deviation from
SM3~\cite{PDG} expectations, we ignore it.

\begin{figure}[t!]
\centerline{{\epsfxsize1.63 in \epsffile{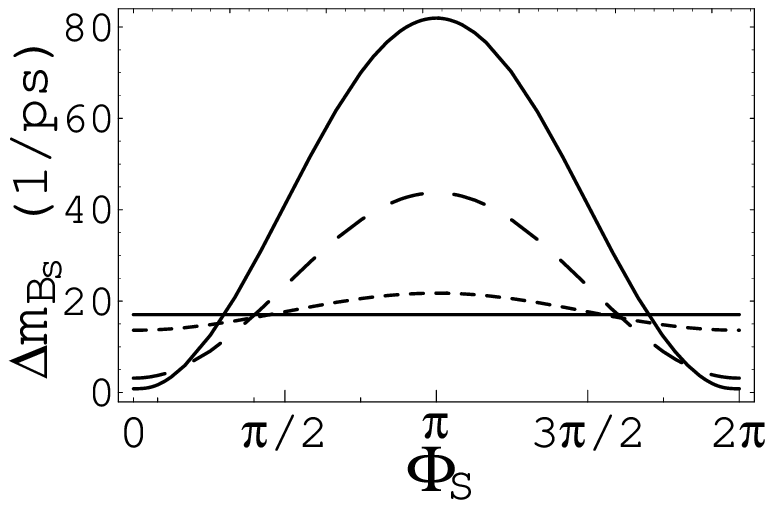}}
            {\epsfxsize1.63 in \epsffile{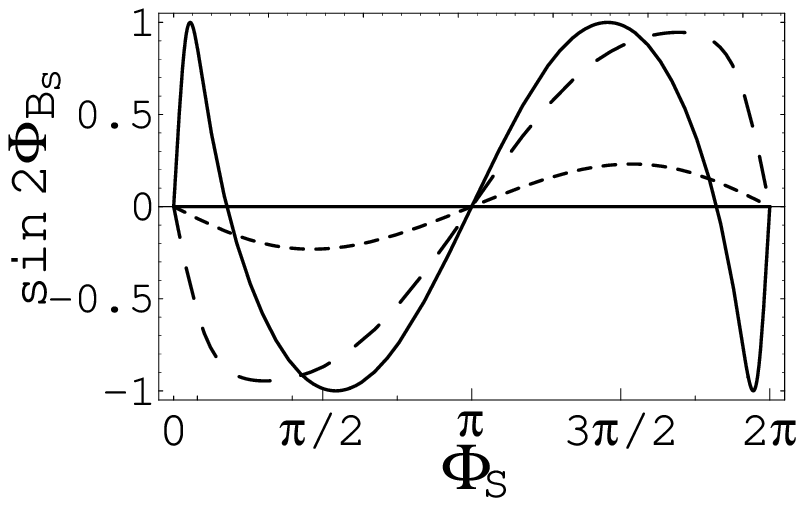}}}
\smallskip
\caption{$\Delta m_{B_s}$ (left) and $\sin 2\Phi_{B_s}$ (right)
vs. $\Phi_s$ for $m_{t^{\prime}}=250$~GeV and $r_s=0.002$ (short
dash), $0.01$ (long dash) and $0.02$ (solid).}
 \label{fig3}
\end{figure}

\begin{figure}[t!]
\centerline{{\epsfxsize1.63 in \epsffile{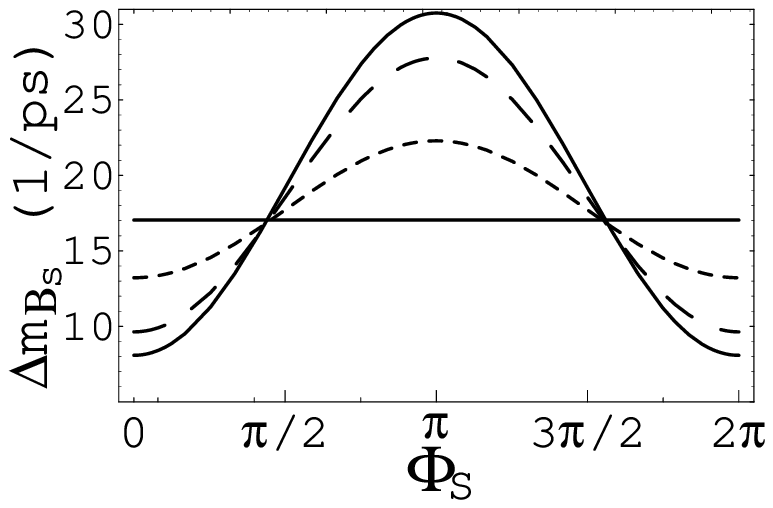}}
            {\epsfxsize1.63 in \epsffile{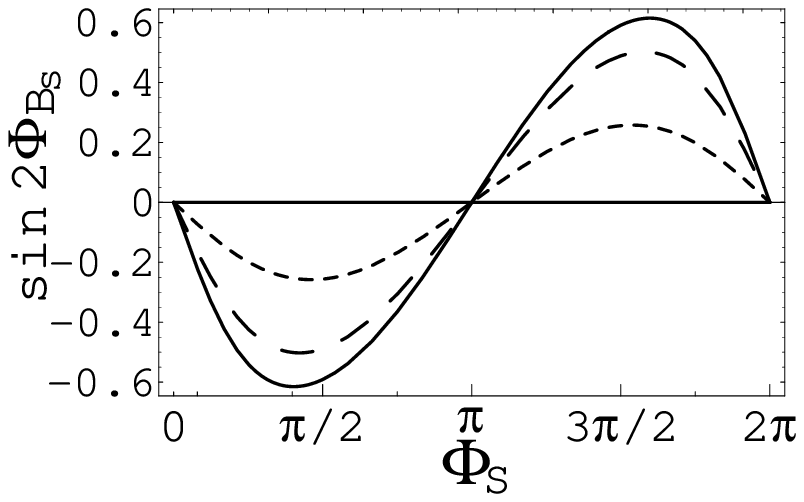}}}
\smallskip
\caption{Same as Fig.~\ref{fig3} for $m_{t^{\prime}}=350$~GeV
and $r_s=0.002$ (short dash), $0.004$ (long dash) and $0.005$ (solid).
} \label{fig4}
\end{figure}

While enhanced $\Delta m_{B_s}$ can easily evade present bounds,
of particular interest is whether and how a fourth generation
could affect the mixing-dependent $CP$ violating observable
$\sin2\Phi_{B_s}$, which is analogous to the recently measured
$CP$ phase in $B_d^0$--$\bar B_d^0$ mixing amplitude.
We plot $\Delta m_{B_s}$ and $\sin 2\Phi_{B_s}$ vs. $\Phi_s$ in
Figs.~\ref{fig3} and \ref{fig4}, for $m_{t'}= 250$ and 350 GeV,
respectively, for several values of $r_s$. In general, the
interference between SM3 and fourth generation effects is
constructive for $\cos\Phi_s <0$, and can enhance $\Delta m_{B_s}$
up to 80--90 ${\rm ps}^{-1}$ for large $r_s\approx 0.02$, as we
have seen already in Fig.~\ref{newfig3}. Conversely, destructive
interference occurs for $\cos\Phi_s>0$, but this would give
$\Delta m_{B_s} < \Delta m_{B_s}^{\rm SM}$ hence disfavored by the
present lower bound. We are thus more interested in the
constructive interference scenario, or when $\Phi_s \sim \pi/2$,
$3\pi/2$, where the $b\to s\gamma$ and $\Delta m_{B_s}$
constraints are more forgiving.

We illustrate the $m_{t'} = $ 250 GeV case with a larger range of $r_s$
values. This not only allows a greatly enhanced $\Delta m_{B_s}$, it makes
$\sin 2\Phi_{B_s}$ very interesting.
We see from Fig.~\ref{fig3} that the $\sin 2\Phi_{B_s}$ structure is rich:
for $r_s>0.01$, $\sin 2\Phi_{B_s}$ can reach $\pm 100$\%.
This can be traced to the dependence of $M_{12}$ on $\Phi_s$.
We see from Eq. (\ref{m12}) that the $\lambda_{t'}^2$, $\lambda_{t'}$, and
$\lambda_{t'}$-indep. terms imply that $M_{12}$ takes the following form:
\begin{eqnarray}
M_{12}=|M_{12}| e^{2 i \Phi_{B_s}} \approx  r_s^2 e^{2 i \Phi_s} A
+ r_s e^{i\Phi_s} B + C
 \label{M12Phis}
\end{eqnarray}
where $A$ and $B$ are explicit functions of $m_t$ and $m_{t'}$ and
$C$ is the usual SM3 contribution. For large enough
$r_s$, the $r_s^2e^{2i\Phi_s}$ term dominates, and one has
$e^{2i\Phi_s}$ modulation. But as $r_s$ decreases, the quadratic
term becomes unimportant, and one is only subject to $e^{i\Phi_s}$
modulation arising from the interference term.

Thus, {\it for $r_s$ as small as 0.002} (compare $\vert
\lambda_c\vert = \vert V_{cs}^* V_{cb}\vert \cong 0.04$), {\it
$\Delta m_{B_s}$ is close to SM3 value, but $\sin 2\Phi_{B_s}$ can
already reach $\mp 25\%$} near $\Phi_s = \pi/2$, $3\pi/2$. For
$r_s=0.01$, $\Phi_s$ is restricted by the $\Delta m_{B_s}$ lower
bound to the range of $70^\circ \lesssim \Phi_s \lesssim
300^\circ$, where $\sin 2\Phi_{B_s}$ can take on {\it extremal
values} of $\mp 100\%$ near the boundaries. This is very
interesting, for {\it if $\Delta m_{B_s}$ is measured soon near
present bounds, we may find the associated $CP$ violating phase to
be {\it maximal}!} For $r_s = 0.02$, the allowed range for
$\Phi_s$ is slightly larger than (roughly $50^\circ \lesssim
\Phi_s \lesssim 310^\circ$) the $r_s=0.01$ case. The maximal $\sin
2\Phi_{B_s}$ now occurs close to $\Phi_s = \pi/2$, $3\pi/2$ since
the fourth generation effect overwhelms the SM3 top quark effect.
In all these cases, $\sin 2\Phi_{B_s}$ vanishes for $\Phi_s = \pi$
when one has maximal constructive interference in $\Delta
m_{B_s}$.

It should be noted that, for purely imaginary $e^{i\Phi_s}$, $r_s$
could be considerably larger than 0.02, leading to very large
$\Delta m_{B_s}$ and maximal $\sin 2\Phi_{B_s}$. The measurement
of such cases, however, are more difficult.

For the heavier $m_{t'} = 350$ GeV scenario of Fig.~\ref{fig4},
$b\to s\gamma$ puts a slightly tighter constraint, hence we have
illustrated the $\Phi_s$ dependence of $\Delta m_{B_s}$ and $\sin
2\Phi_{B_s}$ for $r_s$ only up to 0.005. For this parameter space,
the $r_s^2e^{2i\Phi_s}$ term in Eq.~(\ref{M12Phis}) is
subdominant, and one largely has $e^{i\Phi_s}$ modulation. The
case is similar to the low $r_s$ discussion of Fig.~\ref{fig3}.
Once again, for $\Phi_s \sim \pi/2$, $3\pi/2$, $r_s$ could be much
larger than the illustrated range, and $\sin 2\Phi_{B_s}$ would
become maximal.

\section{Electroweak Penguin $B\to (K,K^*)\ell^+ \ell^-$}

A recent highlight from $B$ factories is the emerging electroweak
penguins. Based on 30 ${\rm fb}^{-1}$ data, the first observation
of $B\to K \mu^+\mu^-$ was reported by the Belle experiment in
2001, giving ${\cal B}(B\to K\mu^+\mu^-) =
(0.99_{-0.32-0.14}^{+0.40+0.13}) \times 10^{-6}$~\cite{belle}.
This was updated~\cite{belle2} in 2002 to
$(0.80_{-0.23}^{+0.28}\pm 0.08) \times 10^{-6}$ with twice the
data at 60~${\rm fb}^{-1}$. Combining with the $B\to Ke^+e^-$
mode, one finds
\begin{eqnarray}
{\cal B}(B\to K\ell^+\ell^-) = (0.58_{-0.15}^{+0.17}\pm 0.06)
\times 10^{-6},
 \label{bkll}
\end{eqnarray}
at 5.4$\sigma$ significance. These results have been
confirmed~\cite{babar2} by the BaBar collaboration with a dataset
of 78~${\rm fb}^{-1}$, giving ${\cal B}(B\to K\ell^+\ell^-) =
(0.78_{-0.20-0.18}^{+0.24+0.11}) \times 10^{-6}$, consistent with
an earlier value of $(0.84_{-0.24-0.18}^{+0.30+0.10}) \times
10^{-6}$ at 56.4~${\rm fb}^{-1}$; the published BaBar upper limit
in 2001, however, gives~\cite{babar} ${\cal B}(B\to K \ell^+
\ell^-)<0.51 \times 10^{-6}$. The latest upper limits for ${\cal
B}(B\to K^* \ell^+ \ell^-)$ are $1.4 \times 10^{-6}$ for Belle
and, for BaBar, $3.0 \times 10^{-6}$, or
$(1.68_{-0.58}^{+0.68}\pm0.28) \times 10^{-6}$ at 2.8$\sigma$
significance.

Given the volatility, including the muon efficiency problem of
BaBar, we shall not combine the above values. However, while
${\cal B}(B\to K\ell^+\ell^-) \sim$ (0.6--0.8) $\times 10^{-6}$ is
not inconsistent with SM predictions~\cite{ali}, it is somewhat on
the high side, as we shall see. Similarly, the new Belle
observation~\cite{belleXsll} of inclusive $B\to X_s\ell^+\ell^-$
decay,
\begin{eqnarray}
{\cal B}(B\to X_s\ell^+\ell^-) = (6.1\pm1.4^{+1.3}_{-1.1}) \times 10^{-6},
 \label{bsll}
\end{eqnarray}
at 5.4$\sigma$ significance,
is also a bit on the high side.

We apply the same approach we have introduced for $b\to
s\gamma$. Using the notations and formulas from
Ref.~\cite{ali}, the effective Hamiltonian for $b\to s \ell^+
\ell^-$ is given by
\begin{eqnarray}
{\cal H}_{\rm eff} = -\frac{4 G_F}{\sqrt{2}} \sum_{i=1}^{i=10} [
\lambda_{t} C_i^{\rm SM}(\mu) + \lambda_{t'} C_i^{\rm new}(\mu) ]
O_i(\mu),
 \label{Hbsll}
\end{eqnarray}
where a similar decomposition as in Eq.~(\ref{decomp}) can be made.
We use the Wilson coefficients $C_i^{\rm SM}$ calculated in naive
dimensional regularization \cite{ndr}. The analytic expressions
for all Wilson coefficients can be found in Ref.~\cite{BBL}.
For simplicity, we will not take into account long-distance effects
from real $c\bar{c}$ intermediate states.

We work with the $\overline{\rm MS}$ bottom mass evaluated at $\mu=5$ GeV.
At the NLO, $m_b(\mu)=m_b^{\rm pole}\{1-\frac{4}{3}
\frac{\alpha_s(\mu)}{\pi}(1-\frac{3}{2}{\log}({m_b^{\rm pole}}/{\mu}))\}$
where $m_b^{\rm pole}=4.8$~GeV is the $b$ quark pole mass.
For other parameters, we take 1.3~GeV as the charm quark mass,
$\overline{m}_t(m_t)=168$ GeV, $m_{K}=0.497$~GeV, $m_{K^*}=0.896$~GeV,
$m_{B}=5.28$~GeV, $|\lambda_c|=|V_{cs}^* V_{cb}|=0.04$ and
$\sin^2\theta_W=0.223$.
To compute the $B\to (K,K^*)\ell^+\ell^-$ rates, we need the $B\to
K$ and $B\to K^*$ transition form factors, where we take the
values from QCD sum rules on the light-cone (LCSR)~\cite{ali}.

We thus obtain the non-resonant branching ratios of ${\cal B}(B\to
K \mu^+ \mu^-)=0.54\times 10^{-6}$ and ${\cal B}(B\to K^* \mu^+
\mu^-)=1.82\times 10^{-6}$ (compare~\cite{ali}) in SM for central
values of the form factors. We note that the {\it more recent NNLO
calculations give results that are 40\% smaller}~\cite{ali2}, and
could suggest a need for new physics, such as the fourth
generation. For $B\to X_s l^+l^-$, with the set of parameters
chosen above, and again without including the long distance
contribution, we find that the non-resonant branching ratio within
SM3 is of order $5.88\times 10^{-6}$, 
where we integrate $\hat{s}=(p_{\ell^+}+p_{\ell^-})^2/m_b^2$ 
over the range $[4 m_l^2/m_b^2,\ 1]$.

\begin{figure}[t!]
\centerline{{\epsfxsize1.71 in \epsffile{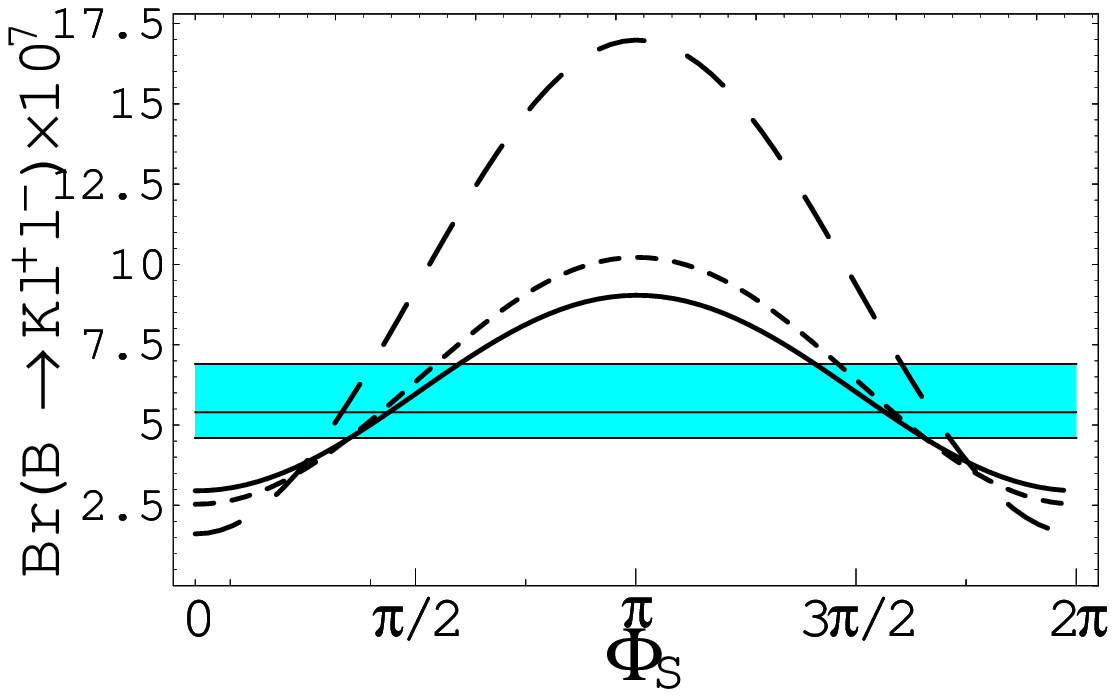}}
            {\epsfxsize1.71 in \epsffile{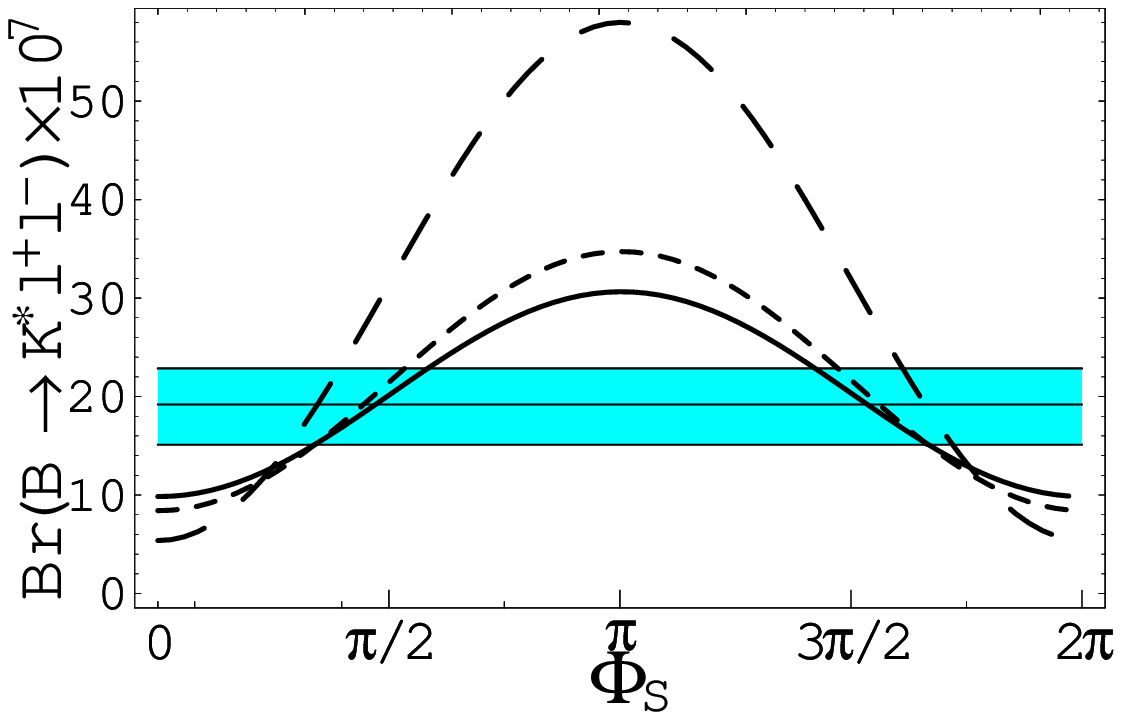}}}
\smallskip
\caption{Non-resonant branching ratios of $B\to K\ell^+\ell^-$ and
$B\to K^*\ell^+\ell^-$, where solid, dashed and long dashed curves
are for $(r_s,\ m_{t'}) = (0.02,\ 250\ {\rm GeV})$, (0.01, 350
GeV), (0.02, 350 GeV), respectively. The horizontal band is the SM
expectation at NLO using LCSR form factors, with central line
standing for central value. The present experimental ranges are
$\sim (0.5$--0.8) $\times 10^{-6}$ and (0.8--1.7) $\times
10^{-6}$, respectively. }
 \label{fig5}
\end{figure}

The strong $m_{t'}$ dependence~\cite{wshou} makes the electroweak
penguin a particularly good probe for the presence of fourth
generation. Fourth generation contributions to $B\to
(K,K^*)\ell^+\ell^-$ have been studied in
Refs.~\cite{wshou,aliev}. These studies, however, took
$\lambda_{t'}$ to be real, which would then be constrained by
$b\to s \gamma$ to two specific $|\lambda_{t'}|$ values for a
given $m_{t'}$~\cite{Huo} (i.e. for $\Phi_s = 0$ and $\pi$).
As we pointed out, allowing for a $CP$ phase in $\lambda_{t'}$, 
for fixed $m_{t'}$ a large region in the
$(\Phi_s,r_s)$ plane is actually still allowed.
In Fig.~\ref{fig5} we plot the non-resonant branching ratio for
$B\to K\ell^+ \ell^-$ and $B\to K^*\ell^+ \ell^-$ vs. $CP$ phase
$\Phi_s$ for several values of $r_s$ and $m_{t'}$.
The horizontal bands give the SM3 range taking into account
form factor uncertainties.
The pattern is similar to $\Delta m_{B_s}$:
for $\cos\Phi_s>0$ we have destructive interference,
and the branching fraction is below SM;
for $\cos\Phi_s<0$ we have constructive interference and the
branching ratio is largest for $\Phi_s=\pi$.

We see that the non-resonant ${\cal B}(B\to K\ell^+\ell^-)$
can reach approximately 2 to 3 times the SM value
depending on $m_{t'}$, $r_s$, and $\Phi_s$. As stated,
such enhancement may become really called for 
if the NNLO result~\cite{ali2} of
${\cal B}(B\to K\ell^+\ell^-) = (0.35\pm0.12)\times 10^{-6}$
is borne out. In any case, a possible factor of 2--3 enhancement
over SM is still within the experimental range of Eq. (\ref{bkll}),
but the case of $r_s\approx 0.02$ with $m_{t'}=350$ GeV may have
too large a branching ratio for $\Phi_s\approx \pi$,
hence $\vert V_{t's}V_{t'b}\vert$ and $m_{t'}$ should not be
simultaneously too large.

\begin{figure}[t!]
\centerline{{\epsfxsize1.63 in \epsffile{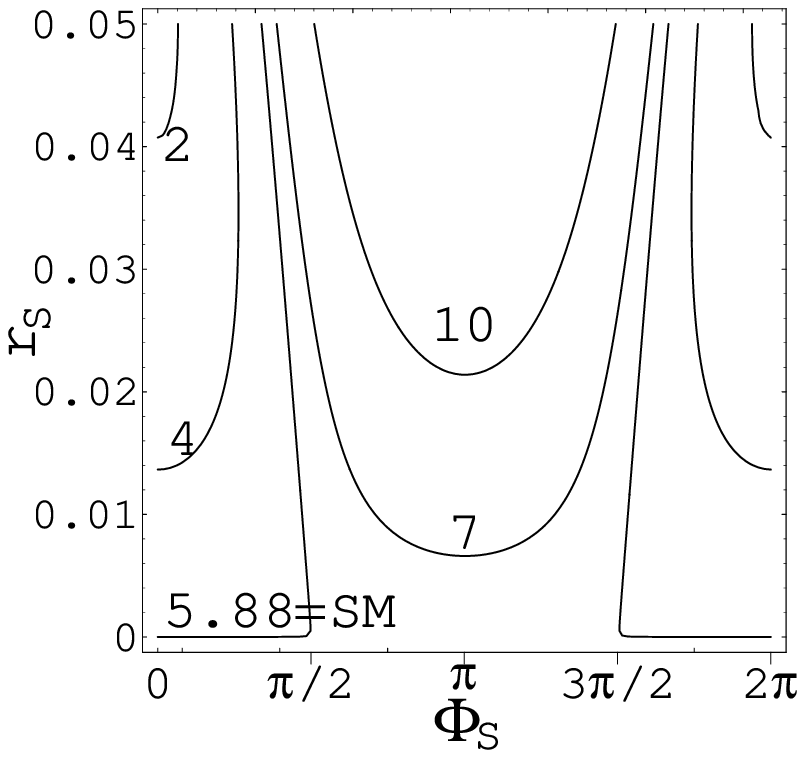}}
            {\epsfxsize1.63 in \epsffile{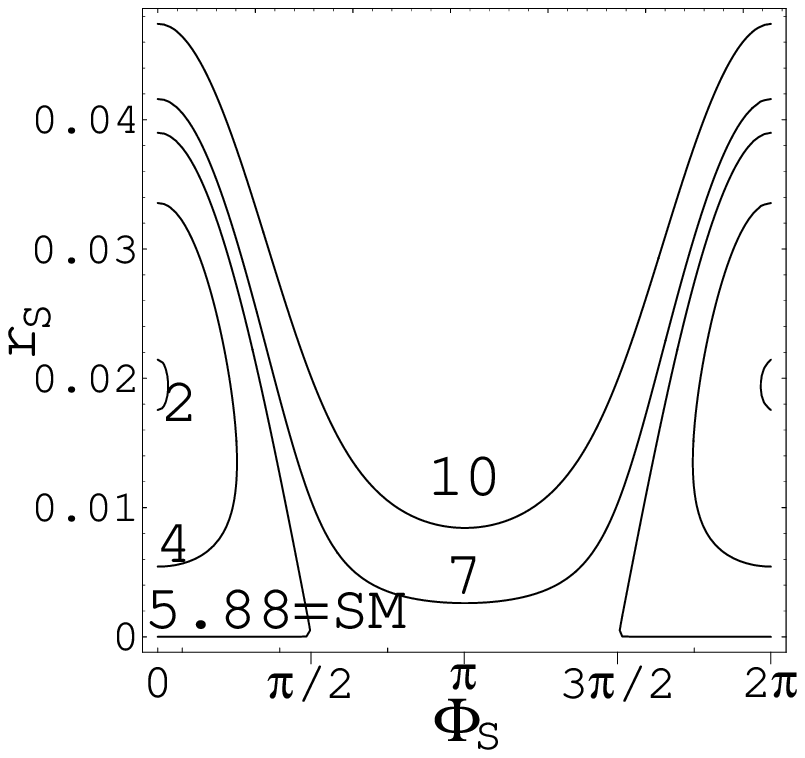}}}
\smallskip
\caption{Contour plot for $B\to X_s\ell^+\ell^-$ in $(\Phi_s,r_s)$ plane
for
$m_{t'}=250\ (350)$ GeV on the left (right).
The present experimental result is
$(6.1\pm1.4^{+1.3}_{-1.1}) \times 10^{-6}$. }
 \label{fig6}
\end{figure}

Thus, the electroweak penguin modes have similar power as the
$\Delta m_{B_s}$ bound, which was illustrated in
Fig.~\ref{newfig3}, in probing fourth generation at present. But
it may improve rapidly once the experiments converge. We note that
there may be some trouble with the ${\cal B}(B\to K^* \ell^+
\ell^-)/{\cal B}(B\to K\ell^+\ell^-)$ ratio between theory and
experiment. This may be either an experimental problem
(fluctuation), or a problem with form factors. For this reason,
the inclusive $B\to X_s\ell^+\ell^-$ may be more interesting,
since it suffers less from hadronic uncertainties. We illustrate
by contour plot in Fig. \ref{fig6} the non-resonant contribution
to ${\cal B}(B\to X_s\ell^+\ell^-) \times 10^{6}$ in the
$(\Phi_s,r_s)$ plane, for $m_{t'}=250$ and 350 GeV. As one can
see, in both cases, $\cos\Phi_s>0$ reduces $B\to X_s l^+l^-$ to be
less than $5.88\times 10^{-6}$ and is less favored, while for
$\pi/2 < \Phi_s < 3\pi/2$ the branching ratio is enhanced and
grows with $r_s$. The behavior is similar to $B_s$ mixing.
We note that the effect of fourth generation on 
$B\to X_s\ell^+\ell^-$ has been studied recently in \cite{yanir}.
The constraints we obtain on $r_s$ from $B\to X_s\ell^+\ell^-$ 
are consistent with Ref.~\cite{yanir}.

The dilepton invariant mass spectrum does not have further
information content beyond the rates themselves.
With proper rescaling, the shape is the same as in SM.
%


Let us now discuss the forward-backward asymmetry (FBA) of $B\to
K^*\ell^+\ell^-$. The usual definition is \cite{ali}
\begin{eqnarray}
\frac{d{\cal A}_{FB}}{d\hat{s}}= -\int_{0}^{\hat{u}(\hat{s})}
d\hat{u} \frac{d^2 \Gamma}{d\hat{u}d\hat{s}} +
\int_{-\hat{u}(\hat{s})}^0
d\hat{u} \frac{d^2 \Gamma}{d\hat{u}d\hat{s}},
\end{eqnarray}
where $\hat{u}(\hat{s})=\sqrt{(1-4{\hat{m}_\ell^2}/{\hat{s}}) \lambda}$ 
and
$\lambda=\lambda(1,\hat{m}_{K^*}^2,\hat{s})=
1+\hat{m}_{K^*}^4+\hat{s}^2 -2 \hat{s} -2\hat{m}_{K^*}^2(1+\hat{s})$
with $\hat{m^2}={m^2}/{M_B^2}$.
From the experimental point of view the normalized FBA
($\bar{\cal A}_{FB}$) is more useful
and is defined by:
\begin{eqnarray}
\frac{d{{\bar{\cal{A}}}}_{FB}}{d\hat{s}}=
\frac{d{\cal A}_{FB}}{d\hat{s}}/\frac{d \Gamma}{d\hat{s}}.
\end{eqnarray}
Note that the FBA for $B\to K\ell^+ \ell^-$ vanishes.

\begin{figure}[t!]
\centerline{{\epsfxsize1.6 in \epsffile{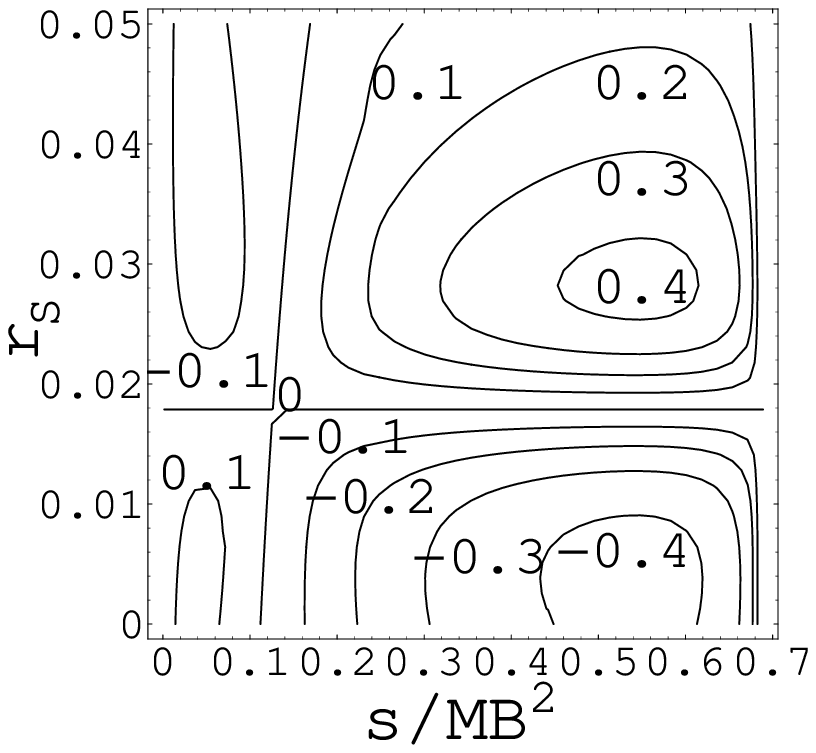}}
            {\epsfxsize1.6 in \epsffile{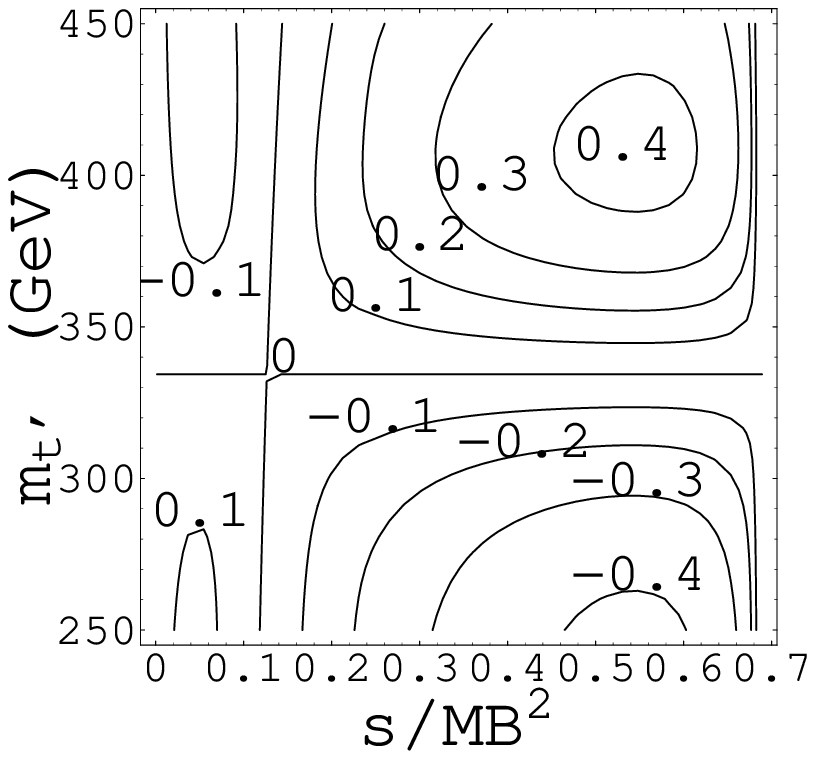}}}
\smallskip
\caption{Normalized forward-backward asymmetry contours for $B\to
K^* \ell^+\ell^-$, for $\Phi_s=0$, in (a) $r_s$ vs. $s/{m_B^2}$
for $m_{t'}=350$ GeV and (b) $m_{t'}$ vs. $s/{m_B^2}$ for
$r_s=0.02$.} \label{fig8}
\end{figure}

In the SM, the FBA for $B\to K^*\ell^+ \ell^-$ is positive for
$\hat{s}<0.1$ with values going up to $\approx 0.1$, vanishes at
$\hat{s}_0=0.1$ and turns negative for $0.7>\hat{s}>0.1$ with FBA
values going down to $\approx -0.4$. This reflects the
interference between the $b\to s\gamma^*$ vs. $b\to sZ^*$ mediated
processes. The presence of a fourth generation affects the latter
more strongly and hence can change the pattern. We are interested
in the parameter space where the shape of FBA change completely
from that of SM. In Fig.~\ref{fig8}, we give contour plots for
normalized FBA, for $\Phi_s = 0$, in the $(\hat{s},\ r_s)$ plane
for $m_{t'}=350$ GeV (left) and in the $(\hat{s},\ m_{t'})$ plane
for $r_s=0.02$ (right). We see that FBA can vanish not only for
$\hat{s}=0.1$ but also in full range of $0.7>\hat{s}>0.1$. In the
left plot this happens for $r_s^0\approx 0.018$ and $m_{t'}=350$
GeV while in the right plot this seems to happen for
$m_{t'}^0\approx 335$ GeV and $r_s=0.02$. From left contour, note
that for $r_s<r_s^0$, the FBA is negative for $\hat{s}>0.1$
exactly as in the SM, but for $r_s
> r_s^0$, the FBA remains positive even for $\hat{s}>0.1$. The
same observation can be made for $\hat{s}>0.1$ in the right
contour: the FBA is negative for $m_{t'}< m_{t'}^0$ but positive
for $m_{t'} > m_{t'}^0$, and FBA can reach +0.4 or more.

\begin{figure}[t!]
\centerline{{\epsfxsize1.6 in \epsffile{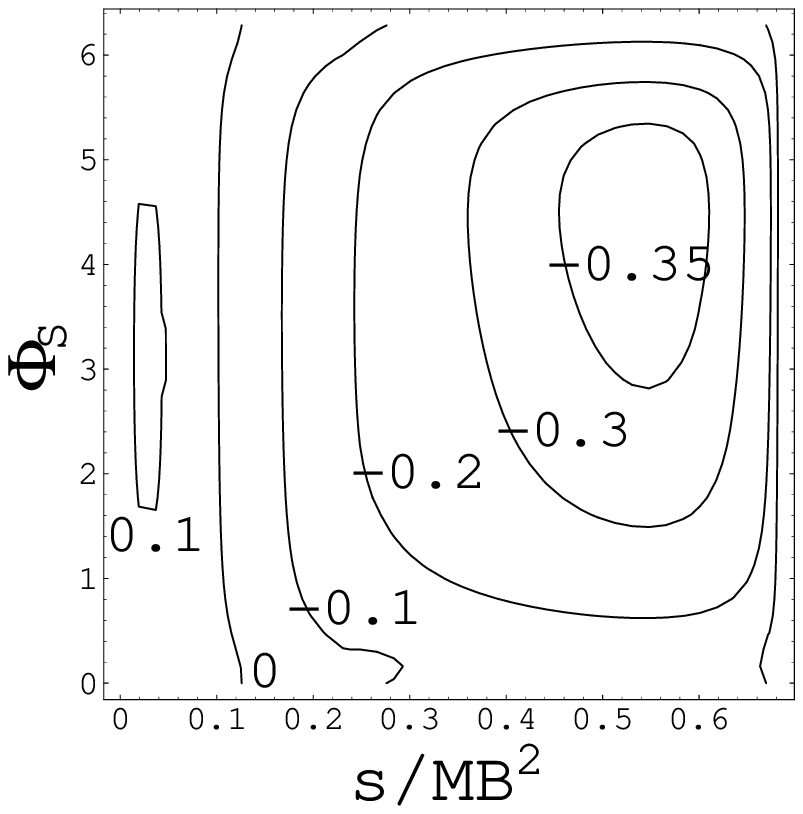}}
            {\epsfxsize1.6 in \epsffile{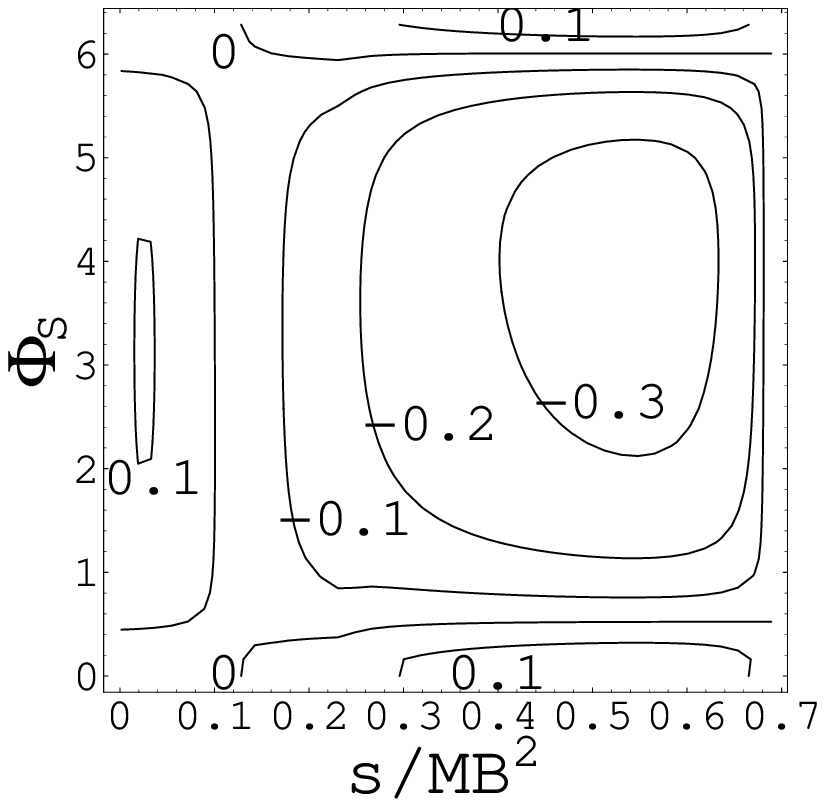}}}
\smallskip
\caption{Normalized $B\to K^* \ell^+\ell^-$ forward-backward
asymmetry contours for $\Phi_s$ vs. $s/{m_B^2}$, for (a)
$m_{t'}=250$ GeV, $r_s=0.04$, and (b) $m_{t'}=350$ GeV,
$r_s=0.02$.} \label{fig9}
\end{figure}

In Fig.~\ref{fig9}, we illustrate the effect of the $CP$ phase
$\Phi_s$ for normalized FBA in the $(\hat{s},\Phi_s)$ plane for
$m_{t'}=250$ GeV, $r_s=0.04$ (left), and for $m_{t'}=350$ GeV,
$r_s=0.02$ (right). From the left plot, one can see that FBA has a
constant sign (negative) for large $\hat{s}>0.1$ and can become
larger than $-0.3$ for large $CP$ phase and large $\hat{s}$. In
the right plot the FBA can flip sign for $\Phi_s\approx 0.5$
and for values of $\hat{s}>0.1$.

\section{$CP$ Violation in Charmless $B$ Decays}

Although it is too early to conclude~\cite{nir}, both Belle and BaBar
presented~\cite{phiKs} some hints for mixing-dependent $CP$
violation in the charmless $B^0\to \phi K_S$ decay mode, with the
two experiments agreeing in sign. It is intriguing that {\it the
sign is opposite that of $B^0\to J/\psi K_S$!} We do not advocate
that new physics is already called for, but illustrate the impact
of a fourth generation.
It is known that the $CP$ asymmetry in both $B^0\to \phi K_S$ and
$J/\psi K_S$ should measure the same quantity $\sin 2 \phi_1$ in
SM, where the uncertainty is estimated to be less than ${\cal
O}(\lambda^2)$ \cite{Gross}.

The effective Hamiltonian for $\Delta B=1$ transition can be written as
\cite{AKL}:
\begin{eqnarray}
{\cal H}_{\rm eff} &= &\frac{G_F}{\sqrt{2}}\left\{ \lambda_u (C_1
O_1^u +C_2 O_2^u)+\lambda_c (C_1 O_1^c +C_2 O_2^c)
\phantom{\sum_{10}^{10}}
\right.\nonumber \\
& &
 \left. \phantom{\sum_{i=1}^{i=1}} -\sum_{i=3}^{i=10} [
\lambda_{t} C_i^{\rm SM}(\mu) + \lambda_{t'} C_i^{\rm new}(\mu) ]
O_i(\mu) \right\}.
 \label{db1}
\end{eqnarray}
For the SM part, we have used the next-to-leading order formulas
given in \cite{AKL} for the Wilson coefficients, while for the
fourth generation effect in $C_i^{\rm new}$, we have used the
leading order analytic formulas given in \cite{BJCF}.

Following the notation of Ref.\cite{AKL}, in the factorization
approach, the amplitude of both $B^-\to K^- \phi$ and
$\bar{B}^0\to \bar{K}^0 \phi$ is proportional to
$V_{tb}V_{ts}^*(a_3+a_4+a_5 -1/2(a_7+a_9+a_{10}))$. The exact form
of the factorizable hadronic matrix element is irrelevant for us
since we are interested only in $CP$ asymmetry observables in SM4,
and the ratio of branching ratios in SM3 and SM4. 
The coefficients $a_i$ are defined as: 
$a_i=C_i^{\rm eff}+1/N_C C_{i\pm1}^{\rm eff}$
for odd, even $i$.

The $CP$ asymmetry $\sin(2 \Phi_{\phi K_S})$ is defined as:
\begin{eqnarray}
\sin(2 \Phi_{\phi K_S}) = -\frac{2\, {\rm Im}\, \Lambda}{1 +
|\Lambda|^2}, \label{sin}
\end{eqnarray}
where $\Lambda=e^{-2i\phi_1} \frac{\overline{{\cal A}}}{{\cal A}}$
with $\overline{{\cal A}}$ being the CP-conjugate amplitude of
${\cal A}$. We define also the following ratio:
\begin{eqnarray}
R_{\phi K_S} =
 \frac{{\cal A}_{\rm SM4}(B^0\to \phi K_S ) }
      {{\cal A}_{\rm SM3}(B^0\to \phi K_S)}.
\end{eqnarray}
In our analysis we use the following set of parameters: $\mu=2.5$
GeV, $m_t=168$ GeV, $m_b=4.88$ GeV,  $k^2=m_b^2/2$, $\alpha=1/128$
and $s_W^2=0.223$ and $\alpha_s(M_W)=0.118$.
In SM3 (i.e. without 4th generation) we find that
$\frac{\overline{{\cal A}}}{{\cal A}}=0.979$ which is close to 1,
and which means that $\sin2\Phi_{\phi K_S} \cong \sin2\phi_1
=\sin2\Phi_{J/\psi K_S}$.
In our numerical evaluation we take $\phi_1$ close to 24$^\circ$
which reproduce the world average of $\sin 2 \phi_1=0.734$ coming
from $B^0\to J/\psi K_S$.

IN SM4, the ratio $\frac{\overline{{\cal A}}}{{\cal A}}$ has
additional $CP$ phase, which we denote as $2\theta$, 
that arises from our $CP$ phase $\Phi_s$. 
So if we use the definition of $\Lambda$ defined above
\begin{eqnarray}
\Lambda  =  e^{-2i\phi_1} \frac{\overline{{\cal A}}}{{\cal A}} 
  = e^{-2i(\phi_1+\theta)} \left\vert {\overline{{\cal A}}\over {\cal
  A}}\right\vert
\end{eqnarray}
where we have used 
$\frac{\overline{{\cal A}}}{{\cal A}}= e^{-i2\theta} 
 |{\overline{{\cal A}}}/{{\cal A}}|$

\begin{figure}[t!]
\centerline{{\epsfxsize1.65 in \epsffile{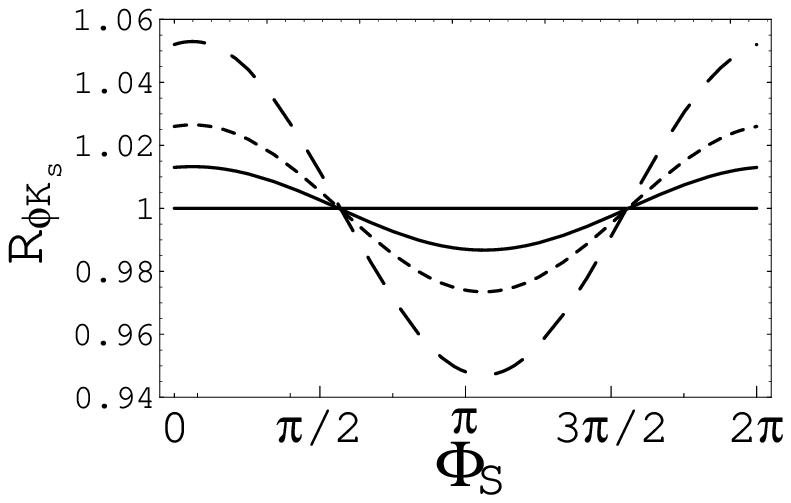}}
            {\epsfxsize1.65 in \epsffile{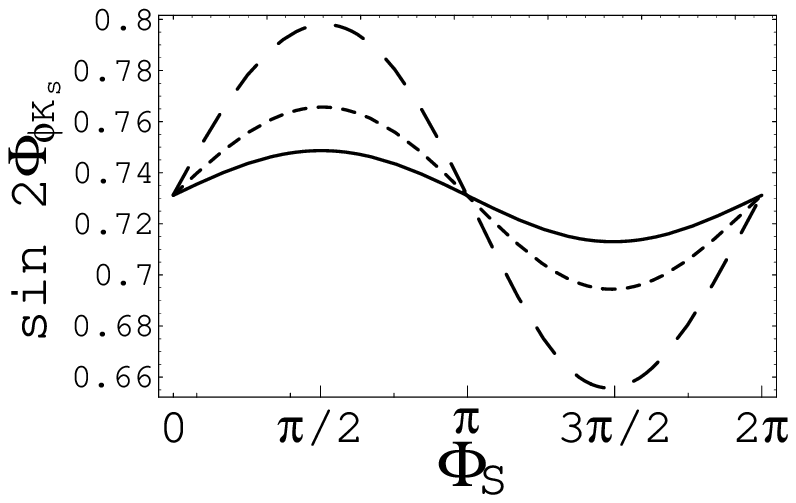}}}
\smallskip
\caption{ (a) Ratio $R_{\phi K_S}$ and (b) $\sin(2 \Phi_{\phi
K_S})$ as function of $CP$ phase $\Phi_s$, where solid, dashed and
long dashed curves are for $r_s$=0.01, 0.02 and 0.04, respectively,
with $m_{t'}=400$ GeV.} \label{fig11}
\end{figure}

We have studied numerically the effect of $CP$ phase $\Phi_s$,
$r_s$ and $m_{t'}$ on both $\sin(2 \Phi_{\phi K_S})$ and $R_{\phi
K_S}$ and found that for $r_s<0.05$ and
$m_{t'}\leq 450$ GeV the effect is rather small. In Fig.
\ref{fig11} we illustrate the effect of $\Phi_s$ on $R_{\phi
K_S}$ (left) and $\sin(2 \Phi_{\phi K_S})$ (right) for
$m_{t'}=400$ GeV and several values for $r_s$. For $r_s=0.02$,
$B^0\to \phi K_S$ in SM4 can be enhanced (reduced) only by about
$2\%$ for $\cos\Phi_s >0$ ($<0$). For large $r_s=0.04$ the effect
can only reach about $5\%$. As one can see in the right plot, the
effect of $CP$ phase $\Phi_s$ on $\sin(2 \Phi_{\phi K_S})$ is also
not impressive. Both in SM3 and SM4, we find numerically that
$a_4\gg a_9/2$ for all $\Phi_s$, while $a_3\gg a_7/2$ and $a_5\gg
a_{10}/2$ in region where $\cos\Phi_s$ is not close to $1$
($\Phi_s\neq 0$ and $2 \pi$). It can be seen that $\sin(2
\Phi_{\phi K_S})$ is enhanced for $0<\Phi_s< \pi$ and reduced for
$\pi <\Phi_s< 2 \pi$. For $r_s=0.02$ ($0.04$) the
enhancement(suppression) is about $+15\%$ ($-30\%$). The
suppression of $\sin(2 \Phi_{\phi K_S})$ is insufficient to drive
it negative, even for rather sizable $r_s$.
For sign change, one would need $\Phi_s \simeq 3\pi/2$ with
unacceptably large $r_s$. 
For example, for $m_{t'} = 400$ GeV and $\Phi_s = 3\pi/2$,
$\sin 2\Phi_{\phi K_S}$ turns negative for $r_s > 0.11$,
which is much larger than $V_{cb}$.

\section{Discussion and Conclusion}

A particularly interesting aspect of the existence of 
a fourth sequential generation is 
the presence of new $CP$ violating phases.
Restricting ourselves to $b\to s$ transitions, 
besides the strength of $\vert V_{t's}V_{t'b} \vert \equiv r_s$,
one has a unique new $CP$ phase, $\arg V_{t's}V_{t'b} \equiv \Phi_s$.
Existing work in the literature have not emphasized 
$\Phi_s \neq 0$, $\pi$ situation, hence have overconstrained the
possible impact of the fourth generation.
In the present work, we have tried to cover as much ground as possible,
and we find a rather enriched phenomenology for $b\to s$ transitions,
for both $CP$ violating or $CP$-indep. observables.

The $b\to s\gamma$ process is not very sensitive to 
the presence of a fourth generation, since the structure is rather
similar to SM, and it is not possible to generate large $CP$ asymmetry
because of the chiral structure. It is therefore especially
accommodating for $\Phi_s = \pi/2$, $3\pi/2$, when fourth generation
contribution adds only in quadrature in rate.
These are the most interesting situations, 
since $CP$ violating effects elsewhere are the {\it largest}.

For example, the $B_s$ mixing and $CP$ phase $\sin 2 \Phi_{B_s}$ are
very interesting probes of fourth generation effects.
While some part of parameter space allowed by $b\to s\gamma$
($\cos\Phi_s \gtrsim 0$) is ruled out, $\Delta m_{B_s}$ can be 
much larger than SM3 expectations.
Any value for $\sin 2 \Phi_{B_s}$ between $-1$ and $+1$ is 
in principle possible.
What may be interesting is that, even for $r_s$ rather small,
hence $\Delta m_{B_s}$ just above present bound,
$\sin 2 \Phi_{B_s}$ could still be sizable.
Thus, Tevatron Run II data is eagerly awaited.

The electroweak penguin modes $B\to K^{(*)}\ell^+\ell^-$ and
$X_s\ell^+\ell^-$ have recently been observed,
and the rates tend to be a little on the high side,
especially when compared to more recent NNLO results.
Again, the best case seems to be for $\Phi_s \sim \pi/2$, $3\pi/2$,
when fourth generation effects add mildly in quadrature.
Although the dilepton spectrum, and roughly speaking 
the direct $CP$ asymmetries as well (analogous to $b\to s\gamma$),
are not much affected,
the forward-backward asymmetry, ${\cal A}_{FB}$, 
can be drastically affected.
This is because the $\gamma^*$ effect is far less sensitive to
fourth generation effects than the $Z^*$ effect, so
the $s = m_{\ell^+\ell^-}^2/m_B^2$ dependence of ${\cal A}_{FB}$ 
is a sensitive probe probe of the interference between the two terms.
Thus, while we need the Belle and BaBar measured rates to converge
better, in the long run the forward-backward asymmetry 
may provide a very interesting probe of the fourth generation.

Finally, we studied the mixing dependent $CP$ violation in the
$B\to \phi K_S$ charmless decay, since some hint for beyond SM 
appeared here recently. We found the effect of fourth generation
to be insignificant here. This is because the strong penguin, 
even more so than $b\to s\gamma$, depends very mildly on 
fourth generation, while the ``harder" $C_9$ term, analogous to
the electroweak penguin contribution to $B\to K^{(*)}\ell\ell$,
is subdominant.
Thus, even with its new source of $CP$ violation,
it is unlikely for the fourth generation 
to change the sign of $\sin 2\Phi_{\phi K_S}$
with respect to $\sin 2\Phi_{J/\psi K_S}$ in the $B_d$ system.

The impact of the fourth generation is most prominent in
$B_s$ mixing and electroweak penguin $b\to s$ transitions,
but mild in electromagnetic and strong penguins.
We have found that the fourth generation, even with 
$\vert V_{t's}V_{t'b} \vert$ not far below 
$\vert V_{cs}V_{cb} \vert \simeq 0.04$,
could be lurking with purely imaginary KM phase.

\begin{acknowledgements}
A. Arhrib is supported by Alexander von Humboldt Foundation.
The work of WSH is supported in part by grant 
NSC-91-2112-M-002-027, the MOE CosPA Project,
and the BCP Topical Program of NCTS.

\end{acknowledgements}

\end{document}